# China's plug-in hybrid electric vehicle transition: an operational carbon perspective

Yanqiao Deng [1], Minda Ma [2 *, 3], Nan Zhou [3], Zhili Ma [1], Ran Yan [1], Xin Ma [4]

1. School of Management Science and Real Estate, Chongqing University, Chongqing, 400045, PR China
2. School of Architecture and Urban Planning, Chongqing University, Chongqing, 400045, PR China
3. Building Technology and Urban Systems Division, Energy Technologies Area, Lawrence Berkeley National Laboratory, Berkeley, CA 94720, United States
4. School of Mathematics and Physics, Southwest University of Science and Technology, Mianyang, 621010, PR China

- Corresponding author: Dr. Minda Ma, Email: maminda@lbl.gov
  Homepage: https://buildings.lbl.gov/people/minda-ma
  http://chongjian.cqu.edu.cn/info/1556/6706.htm

# Graphical Abstract

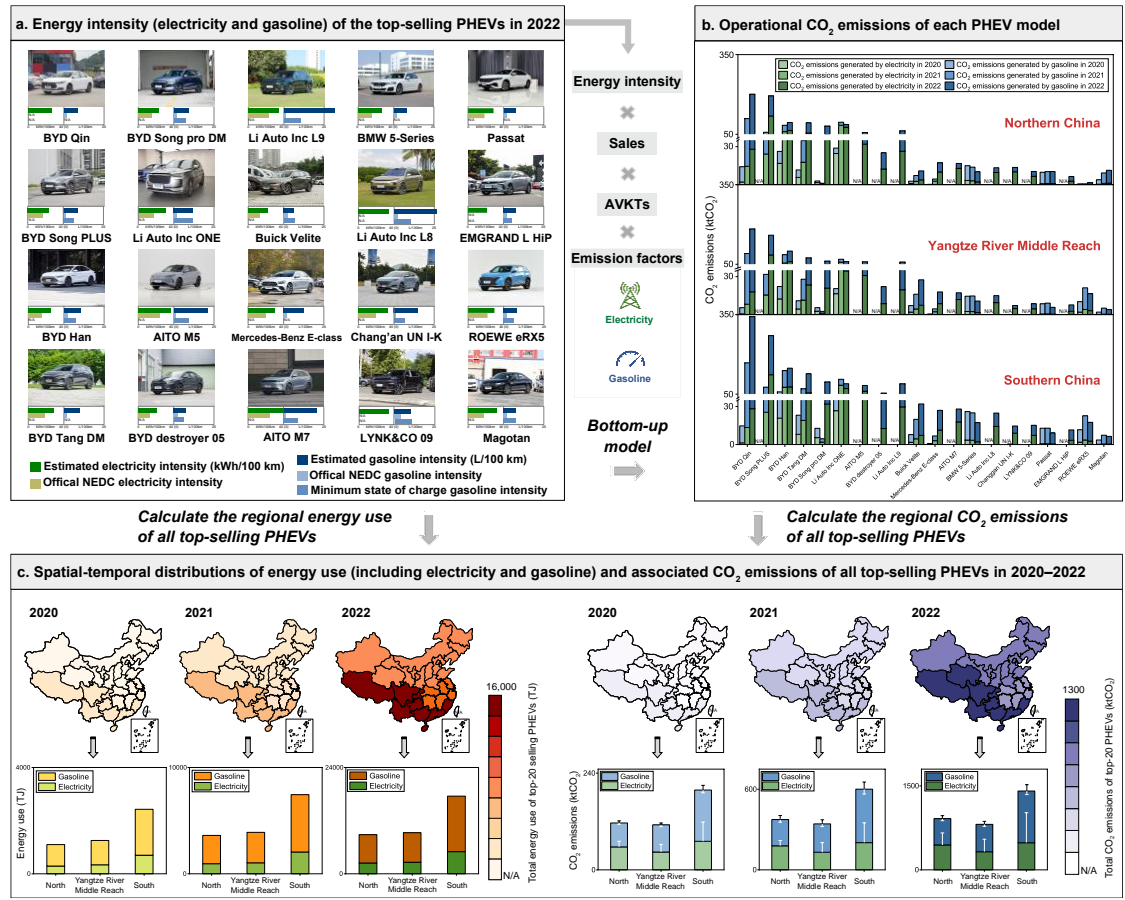

Graphical abstract: (a) Energy intensity (including electricity and gasoline) estimates for the top twenty selling PHEV models in 2022; (b) Operational $CO_2$ emissions of each PHEV model across different regions in China from 2020-2022; (c) Spatial-temporal distributions of energy use (including electricity and gasoline) and associated $CO_2$ emissions for all top-selling PHEV models from 2020-2022.

**Highlights**

- We monitored the operational emissions of China's top-selling PHEVs from 2020-2022.
- The PHEV electricity intensity (0.20-0.38 kWh/km) exceeded its NEDC value by >30%.
- The PHEV gasoline intensity (0.05-0.24 L/km) was 3-6 times higher than its NEDC value.
- The top-selling PHEVs emitted 5 MtCO$_2$ (2 Mt from electricity and 3 Mt from gasoline).
- The energy use of PHEVs in Southern China (27,410 TJ) was double that of other regions.

**Abstract**

Assessing the emissions of plug-in hybrid electric vehicle (PHEV) operations is crucial for accelerating the carbon-neutral transition in the passenger car sector. This study is the first to adopt a bottom-up model to measure the real-world energy use and carbon dioxide emissions of China's top twenty selling PHEV models across different regions from 2020 to 2022. The results indicate that (1) the actual electricity intensity of the best-selling PHEV models (20.2-38.2 kWh/100 km) was 30-40% higher than the New European Driving Cycle values, and the actual gasoline intensity (4.7-23.5 L/100 km) was 3-6 times greater than the New European Driving Cycle values. (2) The overall energy use of the best-selling models varied among different regions, and the energy use from 2020 to 2022 in Southern China was double that Northern China and the Yangtze River Middle Reach. (3) The top-selling models emitted 4.7 megatons of carbon dioxide nationwide from 2020 to 2022, with 1.9 megatons released by electricity consumption and 2.8 megatons released by gasoline combustion. Furthermore, targeted policy implications for expediting the carbon-neutral transition within the passenger car sector are proposed. In essence, this study explores and compares benchmark data at both the national and regional levels, along with performance metrics associated with PHEV operations. The main objective is to aid nationwide decarbonization efforts, focusing on carbon reduction and promoting the rapid transition of road transportation toward a net-zero carbon future.

**Abbreviation notation**

AER – All-electric range

AVKT – Annual vehicle kilometers traveled

BE – Battery energy

CD – Charge-depleting

CS – Charge-sustaining

IPCC – Intergovernmental Panel on Climate Change

$ktCO_2$ – Kilotons of carbon dioxide

LCA – Life cycle analysis

$MtCO_2$ – Megatons of carbon dioxide

NEDC – New European Driving Cycle

PHEV – Plug-in hybrid electric vehicle

SOC – Battery state of charge

TJ – Terajoules

**Nomenclature**

Note: Region $j = S, Y, N, T$, where $S$ represents Southern China, $Y$ represents the Yangtze River Middle Reach, $N$ represents Northern China, $T$ represents the nationwide.

$AVKTE_{i,j}$ – Annual electric vehicle kilometers traveled of vehicle model $i$ in region $j$ (unit: 100 km)

$AVKTG_{i,j}$ – Annual gasoline vehicle kilometers traveled of vehicle model $i$ in region $j$ (unit: 100 km)

$BE_{i,k}$ – Battery energy of the $k$-th vehicle model $i$

$CEE_j$ – $CO_2$ emissions in region $j$ stemming from electricity

$CEG_j$ – $CO_2$ emissions in region $j$ directly generated by gasoline combustion

$EC_{i,j}$ – Total AVKT-based energy consumption of vehicle model $i$ in region $j$

$EI_i$ – Electricity intensity of vehicle model $i$ (unit: kWh/100 km)

$\eta$ – Real-world all-electric range coefficient

$f_{ej}$ – Carbon emission factor of electricity in region $j$

$f_g$ – Carbon emission factor of gasoline

$GI_i$ – Gasoline intensity of vehicle model $i$ (unit: L/100 km)

$\mu_{ei}$, $\mu_{gi}$ – Electricity-to-gasoline ratios of vehicle model $i$

$NAER_{i,k}$ – AER of the $k$-th vehicle model $i$ under the NEDC conditions

$\omega_k$ – Model popularity ratio of the $k$-th vehicle model

$\rho$ – Gasoline density factor

$Sal_{i,j}$ – Sales of vehicle model $i$ in region $j$

$\zeta_e$, $\zeta_g$ – Electricity and gasoline conversion factors, respectively

# 1. Introduction

*1.1. Background*

Sales of electric cars, represented by the plug-in hybrid electric vehicles (PHEVs), nearly doubled year-on-year to 6.6 million globally in 2021 [1], which have the potential to contribute to reducing carbon dioxide ($CO_2$) emissions given the renewable power generation profiles [2, 3]. Although PHEVs are considered eco-friendly and have grown in popularity in recent years, their effects on emission mitigation remain controversial [4]. Several recent studies have shown that real-world $CO_2$ emissions could exceed official emissions [5, 6], as the energy use and corresponding emissions from PHEV operations are sensitive to several potential factors, such as complicated road conditions [7], vehicle weight and speed [8], and individual driving behavior [9]. In particular, the energy use and $CO_2$ emissions of PHEV operations exhibit significant heterogeneity across different regions in China, influenced by ambient temperatures [10] and the power generation mix [11]. To date, few studies have assessed the trends in energy and emissions generated by PHEV operations across various regions, especially in China. Additionally, the various PHEV makes and models prevalent in fierce automotive market competition in recent years were not included.

*1.2. Literature review*

In the field of assessing the energy use and $CO_2$ emissions of PHEV operations, a diverse array of assessment methodologies has gained increasing attention in recent years, mainly focusing on the perspectives offered by simulation methods [12, 13], life cycle analysis (LCA) [14, 15], statistical regression analysis [16, 17], and the application of top-down [18, 19] and bottom-up frameworks [20]. A detailed summary is presented in Table 1 (see the end of the Introduction).

First, a series of energy and emission assessments of PHEVs based on simulation methods have been carried out in recent years, mainly focusing on factors such as average speed [21], battery current and battery state of charge (SOC) [22], charging mode [23], cold start and hot stabilized operation [24], driving behavior and trip condition effects [25], and the type of vehicle and size of the city [26]. However, these studies focused primarily on a limited number of PHEV models from a microscopic perspective and revealed significant variability in methodology, assumptions, data quality, and model design [27, 28]. Additionally, these models are not representative of various PHEVs from different vehicle makes with distinct configurations.

At the national level, studies on PHEV energy and emissions have employed LCA and statistical regression analysis to assess holistic real-world performance across different regions and countries. LCA is a well-established and extensively used systematic tool for comparing the environmental impacts of transportation options across the entire life cycle of a PHEV, including material extraction, manufacturing, transport, use, and end-of-life [29, 30]. The energy and emission estimations using LCA vary greatly depending on location [15, 31], energy mix for electricity generation [32, 33], type of PHEV [34, 35], and driving or charging habits [36]. In terms of statistical regression analysis, empirical studies on the energy and emissions of PHEVs include regression analysis in Canada and the United States (US) [37], quantile-on-quantile regression approaches in eight leading countries [38], and cointegration regression methods for five countries [39]. These studies on regression analysis, as well as LCA, emphasize associations between PHEV adoption and $CO_2$ emissions and have largely dominated the assessments of reduced carbon emissions by PHEVs, which are associated mainly with economic [39, 40] and environmental benefits [41, 42].

The top-down and bottom-up frameworks are effective approaches for assessing the energy and emissions of PHEVs. Hofmann et al. [43] developed a top-down framework to assess the reduction in $CO_2$ emissions from electric vehicles in China at a nationwide scale. However, the utilization of a top-down approach that relies on annual data to explore the interplay between PHEVs and $CO_2$ emissions introduces notable bias in emission estimates [44]. This bias arises from a lack of detailed information, with a disproportionate emphasis on observed macroeconomic trends. Conversely, the bottom-up approach tends to use micro input data to construct a more systematic energy and emission estimation model from the ground up, starting with detailed information such as vehicle makes/models and fuel types, in China's road transport sector. For example, Lu et al. [45] developed a bottom-up approach to measure the $CO_2$ emissions of high-frequency passenger car sales data from 2016-2019 in China, effectively reducing the uncertainty of carbon emission accounting.

*1.3. Motivation, contributions, and the organization*

Regarding the assessment of the energy and emissions of PHEV operations, the macroscopic approach involves estimating $CO_2$ emissions using annual data from national statistical yearbooks, resulting in biased estimates without considering specific technical details of PHEVs [38, 39, 46]. The microscopic approach focuses primarily on carbon intensity but neglects total PHEV sales and the annual vehicle kilometers traveled (AVKT), leading to an incomplete overview of total emissions in spatial-temporal scopes [47]. Importantly, these studies focus on several types of PHEV models [9, 28], as real-world simulations under complicated road conditions for different PHEV models are costly and impractical. To estimate the real-world energy and emissions of PHEVs, real-world data should be incorporated into assessment models [48]. However, few existing works have systematically estimated the real-world $CO_2$ emissions of more than twenty PHEV models with various configurations across different regions. To address these gaps, this study proposes the following three issues for top-selling PHEVs in the passenger car sector of China:

- How can a real-world end-use energy model be established for top-tier PHEV makes and models?
- What is the heterogeneity in the energy use of PHEV operations across different regions?
- How can the operational carbon trends of PHEVs be measured at the national and regional scales?

**To make the most significant contributions,** this study pioneers the development of a standardized bottom-up end-use framework specifically tailored to assess the energy (electricity and gasoline) use and corresponding emissions of PHEV operations across three regions—Southern China, the Yangtze River Middle Reach, and Northern China—from 2020-2022. By establishing a robust foundation of credible data from current PHEV energy use, this study sets a baseline. This baseline not only enables the modeling of potential emissions and energy for PHEV operations in the coming years but also serves as a valuable tool to decarbonize passenger cars up to 2060. This work evaluates the spatial-temporal transition features of both the intensity and total of emissions and energy of PHEVs across different regions in China. This effort is aimed at expediting the transportation sector's move toward carbon neutrality.

The rest of this paper proceeds as follows: Section 2 introduces the developed bottom-up model and the datasets used. Section 3 provides the estimated energy and emissions of PHEVs among different regions. Section 4 further discusses the comparison results, the uncertainty and sensitivity analyses, and the policy implications. Section 5 summarizes the core findings and proposes future studies.

Table 1. Summary of major studies assessing the energy and emissions of PHEVs since 2019.

| Methodology | Assessment techniques | Vehicle types | Case areas | Reference |
|---|---|---|---|---|
| Simulation | Real-world usage data | PHEVs (samples: approximately 100,000 vehicles) | China, Germany, Norway, US, Canada, and Netherlands | Plötz et al. [5] |
| | Laboratory and on-road tests | PHEVs (samples: 4 models) | Europe | Tansini et al. [6] |
| | Real-driving emissions test | PHEVs (samples: 3 models) | Germany | Ehrenberger et al. [9] |
| | Calibrated vehicle simulators | PHEVs (samples: 2 models) | Brussels and Belgium | Dauphin et al. [12] |
| | Empirical study | PHEVs (samples: 10,488 Chevrolet Volt vehicles) | US and Canada | Mandev et al. [37] |
| Life cycle analysis | Life cycle assessment | PHEVs, BEVs, and ICEVs | China | Lu et al. [34] |
| | Life cycle assessment | PHEVs, BEVs, and ICEVs | China | Peng et al. [35] |
| Statistical regression analysis | Panel data, fixed effects | PHEVs and HEVs | US (state-level) | Squalli [4] |
| | Artificial intelligence model | PHEVs and BEVs | Xi'an city, China | Zhao et al. [26] |
| Bottom-up approach | A combinational optimization assessment model | PHEVs and BEVs | China | Wang et al. [20] |
| | A low-carbon transition planning model | EVs | China | Lu et al. [45] |

Note: PHEVs denote the plug-in hybrid vehicles, BEVs denote the battery electric vehicles, ICEVs denote the internal combustion engine vehicles, HEVs denote the hybrid electric vehicles, and EVs denote the electric vehicles (including all types of electric vehicles, such as BEVs, PHEVs, and HEVs).

## 2. Methods and materials

This work developed a bottom-up estimation framework to measure the energy use and corresponding $CO_2$ emissions of top-selling PHEV model operations in various regions of China. Section 2.1 introduces the bottom-up energy consumption model for PHEVs adaptive to three regions, incorporating a real-world energy intensity estimation that considers comprehensive road conditions and diverse vehicle models. Section 2.2 develops a $CO_2$ emission estimation model that separately considers electricity and gasoline consumption for PHEV operations. Finally, Section 2.3 outlines the datasets and parameter assumptions used in this work.

*2.1. Bottom-up energy consumption assessment for vehicle operations*

The annual total energy consumption of the top-$n$ PHEV sales models in region $j$ (including Southern China, the Yangtze River Middle Reach, Northern China, and nationwide, abbreviated as $EC_S, EC_Y, EC_N, EC_T$, respectively) is estimated via Eq. 1:

$$EC_j = \sum_{i=1}^{n} EC_{i,j} \times Sal_{i,j}, \quad (j = S, Y, N, T) \tag{1}$$

where $EC_{i,j}$ represents the total AVKT-based energy consumption of vehicle model $i$ in region $j$ and where $Sal_{i,j}$ represents the sales of vehicle model $i$ in region $j$ based on the National New Vehicle Compulsory Traffic Insurance.

Given the assumption of charge-depleting (CD) mode priority when the PHEV is fully charged and charge-sustaining (CS) mode or blended modes when the SOC reaches the lowest values, $EC_{i,j}$ [45] can be formulated via Eq. 2:

$$EC_{i,j} = EI_i \times AVKTE_{i,j} \times \zeta_e + GI_i \times AVKTG_{i,j} \times \rho \times \zeta_g \tag{2}$$

where $EI_i$ represents the electricity intensity (unit: kWh/100 km) of vehicle model $i$, with only the electric engine propelling the vehicle, and where $GI_i$ is the gasoline intensity (unit: L/100 km) of vehicle model $i$ under comprehensive real-world road conditions. $AVKTE_{i,j}$ and $AVKTG_{i,j}$ (unit: 100 km) denote the annual electric and gasoline vehicle kilometers traveled by vehicle model $i$ in region $j$, respectively. To unify the units of energy use, the electricity conversion factor $\zeta_e$ is used for converting electricity consumption from kWh to Terajoules (TJ)[a]. $\rho$ is the density factor, which converts the gasoline volume from L to kg[b]. Then, the gasoline conversion factor $\zeta_g$ is used for converting gasoline consumption from kg to TJ.

For estimating $EI_i$ and $GI_i$, there are diverse models with distinct vehicle configurations (i.e., battery energy (*BE*), all-electric range (*AER*), 0-100 km/h acceleration and other features), and they exhibit different levels of popularity in real-world sales and usage. Therefore, an average $EI_i$ [49] is estimated considering the diversity of vehicle models and it can be formulated via Eq. 3:

$$EI_i = \sum_{k=1}^{m} \omega_k \frac{BE_{i,k}}{\eta NAER_{i,k}} \times 100 \quad (i = 1,2,\ldots,n) \tag{3}$$

where $\omega_k$ represents the model popularity ratio of the $k$-th vehicle model, which is determined by the ratio of the actual users of the $k$-th model to the total users of all $i$ models with various configurations, and where $m$ is the number of all the vehicle models $i$ on sale in that year. $BE_{i,k}$ denotes the battery energy of the $k$-th vehicle model, and $NAER_{i,k}$ is the AER under the New European Driving Cycle (NEDC) conditions; that is, the car relies only on the power in the battery to support the maximum driving range of the vehicle in CD mode. The real-world all-electric range coefficient $\eta$ is considered in this work because the real-world AER is often shorter than the official AER under the NEDC conditions [50].

For $GI_i$, the gasoline consumption under the NEDC conditions significantly deviates from that in the real-world situation. Additionally, information on gasoline consumption when the battery SOC reaches its lowest value is incomplete for most vehicle models, especially in blended modes. Therefore, the average $GI_i$ of vehicle model $i$ is estimated via Eq. 4:

$$GI_i = \sum_{k=1}^{m} \omega_k GI_{i,k}, \quad (i = 1,2,\ldots,n) \tag{4}$$

where $GI_{i,k}$ is the real-world comprehensive gasoline consumption per 100 km of the $k$-th vehicle model belonging to vehicle type $i$, as publicly measured by the users of the BearOil app under comprehensive road conditions, which can relatively reflect the actual comprehensive gasoline consumption level.

---

[a] kWh denotes kilowatt-hour (the energy delivered by one kilowatt of power for one hour), which is a non-SI unit of energy equal to 3.6×10⁻⁶ terajoules in SI units.

[b] The SI symbol L denotes liter (a metric unit of volume), and kg denotes kilogram (the base unit of mass). One liter of gasoline has a mass of approximately 0.74 kilograms.

To estimate the corresponding $AVKT$ of electricity consumption and gasoline consumption of vehicle model $i$ in region $j$, $AVKTE_{i,j}$ and $AVKTG_{i,j}$ can be formulated via Eq. 5:

$$\begin{aligned} AVKTE_{i,j} &= \mu_{ei} \times AVKT_j \\ AVKTG_{i,j} &= \mu_{gi} \times AVKT_j \end{aligned} \tag{5}$$

where $\mu_{ei}$ and $\mu_{gi}$ represent the electricity-to-gasoline ratios, which are defined as the ratio of cumulative electricity consumption to cumulative fuel consumption for all samples in all vehicle models $i$ with different configurations.

*2.2. Bottom-up carbon emission assessment for vehicle operations*

The $CO_2$ emissions released by PHEV operations are distinct from those released by the consumption of electricity and gasoline. It is imperative to estimate these components separately, avoiding reliance on a comprehensive energy consumption approach, as there are significant variations in the mechanisms of emissions generation and the carbon emission factors between electricity and gasoline. Consequently, the $CO_2$ emissions in region $j$ (abbreviated $CE_j$) [51] are calculated by Eq. 6:

$$CE_j = CEE_j + CEG_j, \quad (j = N, Y, S, T) \tag{6}$$

where $CEE_j$ and $CEG_j$ represent the $CO_2$ emissions generated by electricity and gasoline, respectively, during the operation of top-selling PHEVs, and $CEE_j$ is calculated by Eq. 7:

$$CEE_j = f_{ej} \times \sum_{i=1}^{n} \left( EI_i \times AVKTE_{i,j} \times Sal_{i,j} \right) \tag{7}$$

where $f_{ej}$ is the carbon emission factor of electricity in region $j$. In addition, $CEG_j$ is formulated via Eq. 8:

$$CEG_j = f_g \times \sum_{i=1}^{n} \left( GI_i \times AVKTG_{i,j} \times Sal_{i,j} \right) \times \rho \times \zeta_f \tag{8}$$

where $f_g$ is the carbon emission factor of gasoline.

*2.3. Datasets*

To evaluate the energy and emissions in China's PHEV operations, the top twenty selling PHEV models across Southern China, the Yangtze River Middle Reach, and Northern China from 2020-2022 were selected. The top twenty selling PHEV models, including the BYD Qin, BYD Song PLUS, BYD Han, BYD Tang DM, BYD Song pro DM, Li Auto Inc. ONE, AITO M5, BYD destroyer 05, Li Auto Inc. L9, Buick Velite, Mercedes-Benz E-class, AITO M7, BMW 5-Series, Li Auto Inc. L8, Chang'an UN I-K, LYNK&CO 09, Passat, EMGRAND L HiP, ROEWE eRX5, and Magotan, were accessed from Autohome (https://www.autohome.com.cn/), along with annual sales data based on the National New Vehicle Compulsory Traffic Insurance. Notably, extended-range EVs, such as the Li Auto Inc. ONE, L8, and L9, as well as AITO M5 and M7 included in this study, were also classified as PHEV models in this study. Additionally, data on BE and AER under the NEDC conditions, which were used to estimate real-world electricity intensity, were also sourced from Autohome, and official data on NEDC electricity consumption, NEDC comprehensive gasoline consumption, and minimum charging state fuel consumption were also included for comparative analysis. To estimate the real-world gasoline intensity of each PHEV model operation, the real-world gasoline consumption per 100 km was sourced from the BearOil app (https://www.xiaoxiongyouhao.com/, accessed in July 2024) [45]. Additionally, the AVKTs in this study were based on the work of Ou et al. [52] from Oak Ridge National Laboratory.

For the model parameters in this study, the real-world all-electric range coefficient $\eta$ was assumed to be 75%, as detailed in the research conducted by Plötz et al. [53]. The model popularity ratios of the $k$-th vehicle model $\omega_k$ and the electricity-to-gasoline ratios $\mu_{ei}$ and $\mu_{gi}$ of each PHEV model were also collected from the BearOil app. Additionally, the carbon emission factors of electricity $f_{ei}$ (unit: kgCO$_2$/kWh) for specific regions were obtained from the study conducted by Zhuo et al. [54], and the operating margin electricity carbon emission factors, released by the Ministry of Ecology and Environment of the PR China (https://www.mee.gov.cn/) and the National Center for Climate Change Strategy and International Cooperation (http://www.ncsc.org.cn/), were collected for uncertainty analysis. The gasoline carbon emission factors $f_g$ (unit: kgCO$_2$/TJ) were sourced from the Intergovernmental Panel on Climate Change (IPCC), including the lower, default, and upper values. More details can be found in the report from the IPCC [51].

## 3. Results

This section presents the results of the energy intensity, total energy use, and corresponding $CO_2$ emissions of top-selling PHEV operations from 2020-2022 across various regions in China. Section 3.1 provides estimates of the electricity and gasoline intensity for the top twenty selling PHEV models. Section 3.2 explores the spatial-temporal distribution characteristics and regional differences in the energy use of PHEV operations. Section 3.3 delves into a comparative analysis of the $CO_2$ emissions generated from electricity and gasoline for each PHEV across different regions and nationwide and further studies the spatial-temporal distributions and regional differences of the corresponding $CO_2$ emissions in PHEV operations.

*3.1. Energy intensity of the top-selling vehicle operations*

Fig. 1 shows an overview of the electricity intensity over the top twenty selling PHEV models. The estimated electricity intensity, characterized by electricity consumption per 100 km, varied significantly among models with different BE and AER configurations, ranging from the most electricity-efficient BYD Destroyer 05 at 20.2 kWh/100 km to the higher-intensity AITO M7 at 38.2 kWh/100 km. The estimated electricity intensity of the most popular PHEV models in China ranged from 21.1-31.5 kWh/100 km. Compact sedans such as the BYD Destroyer 05 and EMGRAND L HiP were notable for their low electricity intensity, with values of 20.2 and 20.7 kWh/100 km, respectively. In contrast, PHEV models developed from internal combustion engines, such as the Mercedes-Benz E-class, Magotan, and LYNK&CO 09, exhibited relatively high electricity intensities, ranging from 29.0-31.5 kWh/100 km. Extended-range EVs with large BE and long AERs, including the Li Auto Inc. ONE, L9, and L8, as well as the AITO M5 and M7, presented the highest electricity intensity levels, ranging from 29.6-38.2 kWh/100 km. In terms of time periods, the electricity intensity of models such as the BYD Song pro DM, BYD Han, BMW 5-Series, BYD Tang DM, and Mercedes-Benz E-class improved after 2020, with reductions ranging from 0.9-8.8 kWh/100 km, reflecting advancements in electric drive train technology. Newly released PHEV models in 2022, including the BYD Destroyer 05, EMGAND L HiP, Li Auto Inc. L8 and L9, and AITO M5 and M7, showed varied electricity intensities. The BYD destroyer 05 and EMGAND L

HiP models presented the lowest electricity intensity, whereas the other models presented the highest electricity intensity.

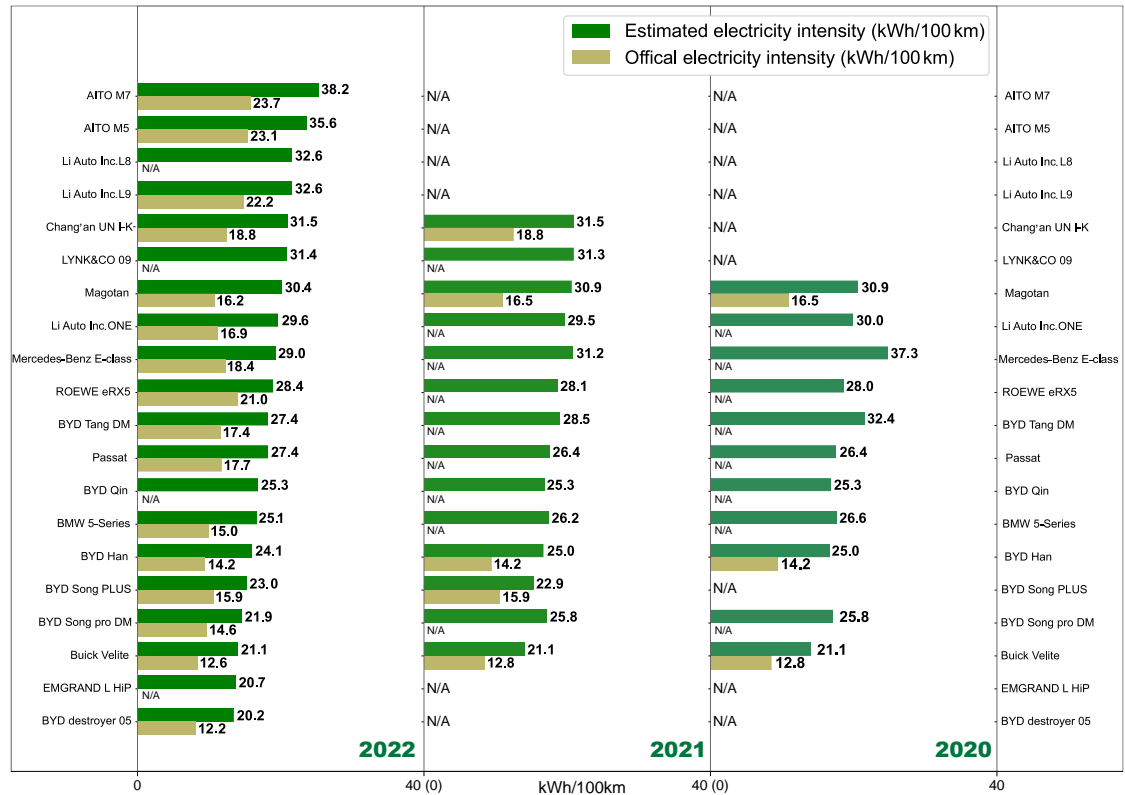

**Fig. 1.** Comparison of estimated and official operational electricity intensities for China's top twenty selling PHEV models from 2020-2022. Note: The green bars represent the estimated electricity intensity, whereas the khaki bars represent the official electricity intensity under the NEDC conditions.

Considering the electricity intensity associated with the estimated average BE and real-world AER of the top twenty selling PHEV models, as shown in Fig. 2, models with lower BE and AER tended to exhibit lower electricity intensity levels. Conversely, as the BE and ARE increased, so did the electricity intensity. This correlation aligned with expectations for most models, where a larger battery capacity facilitates greater energy storage, potentially extending the electric driving ranges and, consequently, resulting in relatively higher electricity intensity levels [55]. For example, the specific extended-range EVs of AITO resulted in the highest average energy intensity at 36.9 kWh/100 km, with an average BE of 40 kWh and an average AER of 108.7 km. The models of Li Auto Inc. reached an average energy intensity of 31.6 kWh/100 km, with an average AER reaching 131.4 km. However, PHEV models developed from traditional internal combustion engines with lower BE and AER, including the ROEWE eRX5, Passat, and Magotan, exhibited considerably greater energy intensities than did other PHEV models and were even comparable to the Li Auto Inc. ONE model. Furthermore, 60% of the top-selling PHEV models have BEs ranging from 10-20 kWh to maintain moderate electricity intensity. However, these PHEV models are not favored because of concerns about being "not electric enough" for a long driving period [37]. As a result, real-world electricity intensity estimates reveal that most current PHEV models need to optimize electricity powertrains or energy management systems to meet the slogan of being "efficient electric vehicles" [56].

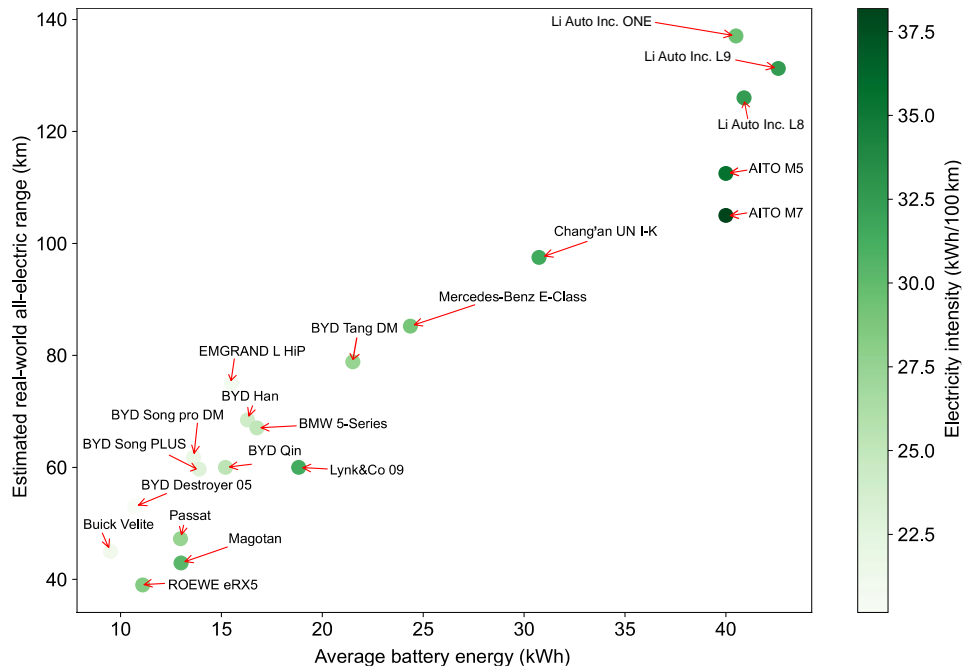

**Fig. 2.** Operational electricity intensity associated with the average BE and estimated real-world AER of the top twenty selling PHEV models in China (2022).

Regarding the gasoline intensity analysis, Fig. 3 provides an overview of the gasoline intensity estimates among the top twenty selling PHEV models in China. Overall, the estimated real-world gasoline intensity, measured by gasoline consumption per 100 km, varied significantly among different PHEV models, from the model with the best fuel economy—the Buick Velite—at 4.7 L/100 km to the model with the highest fuel consumption—the Li Auto Inc. L9—at 23.5 L/100 km. The gasoline intensities of the top twenty selling PHEV models were distributed at three levels. The most efficient level, ranging from 4.7-5.9 L/100 km, included models such as the Buick Velite, Passat, Magotan, BYD Destroyer 05, BMW 5-Series, and EMGRAND L HiP. These models exemplified fuel-efficient internal combustion engines, closely aligning with the ideal gasoline intensity scenario envisioned for PHEV development in China. The moderate level, spanning from 6.0-7.5 L/100 km, encompassed models such as the ROEWE eRX5, BYD Qin, BYD Song Pro DM, BYD Song PLUS, BYD Tang DM, and LYNK&CO 09, reflecting the acknowledged gasoline intensity in real-world PHEV development. Finally, the high gasoline intensity level, ranging from 8.9-23.5 L/100 km, included models such as the Li Auto Inc. ONE, Mercedes-Benz E-Class, Chang'an UNI-K, AITO M5, AITO M7, Li Auto Inc. L8, and Li Auto Inc. L9, which appeared more akin to internal combustion engines than fuel-efficient PHEVs. Notably, specific extended-range EVs, including the AITO M5 and M7 and the Li Auto Inc. L8 and L9 (excluding the Li Auto Inc. ONE), had the highest gasoline intensity levels, ranging from 15.5-23.5 L/100 km, exceeding those of some internal combustion engine vehicles. Over the period from 2020-2022, the gasoline intensities of most PHEV models remained relatively stable, except for the BYD Tang DM, with a 1.8 L/100 km decrease, and the BYD Song pro DM, with a 0.8 L/100 km decrease. Overall, the results above portray the energy intensity of the operation of China's top-selling PHEV models and answer Issue 1 posed in Section 1.

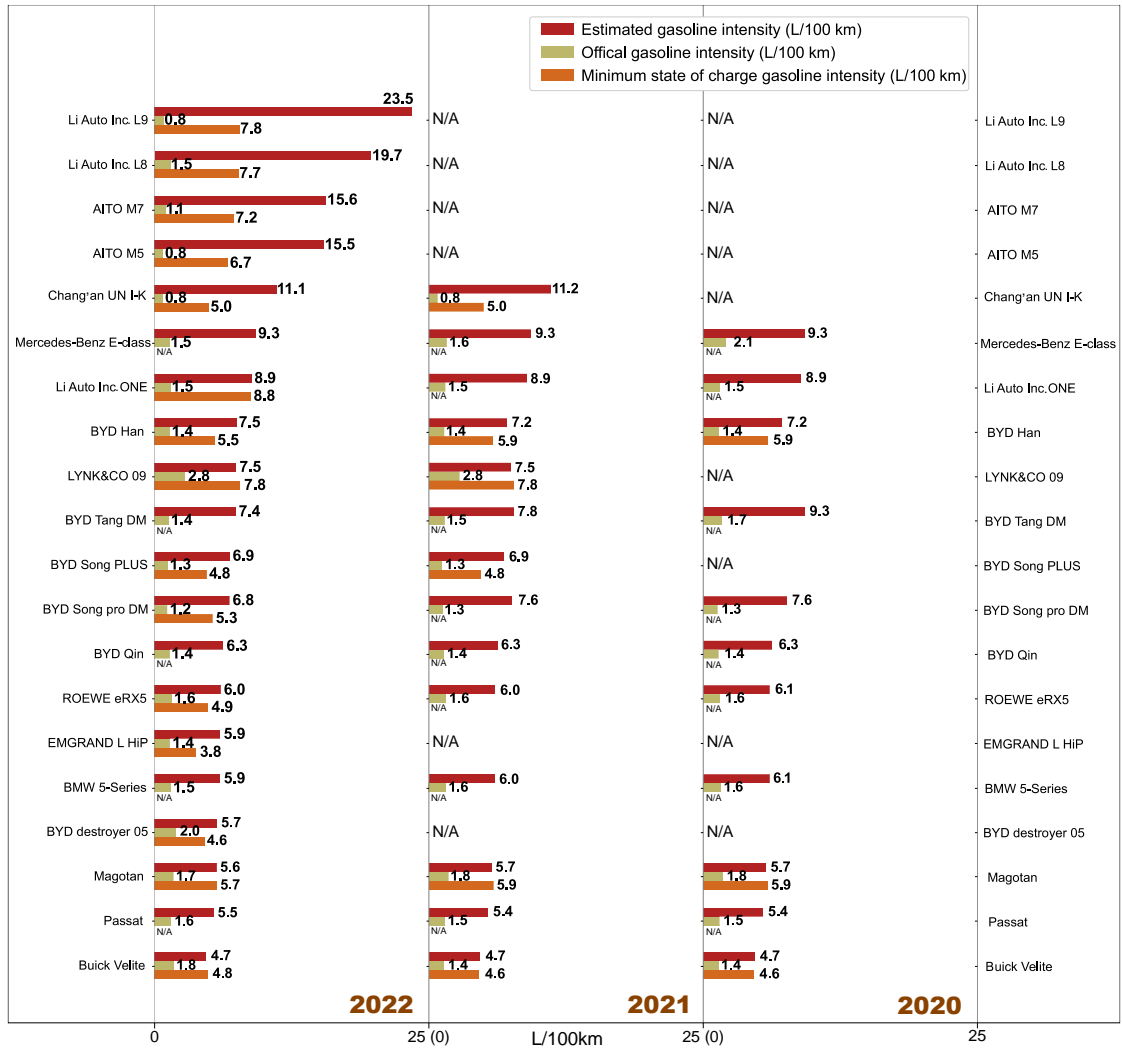

**Fig. 3.** Comparison of estimated and official gasoline use intensities for China's top twenty selling PHEV models from 2020-2022. Note: the red bars represent the estimated gasoline consumption intensity, the yellow bars represent the official gasoline intensity under the NEDC conditions, and the orange bars represent the minimum state of charge gasoline consumption intensity.

*3.2. Operational energy use of the top-selling vehicles*

Fig. 4 presents the operational energy consumption estimates of the top twenty selling PHEV models from 2020-2022 across the regions of Southern China, the Yangtze River Middle Reach, and Northern China, along with vehicle sales in Fig. 4 a; comparisons of the energy consumption estimated by the proposed bottom-up approach and the BearOil app in Fig. 4 b; and the corresponding AVKTs powered by electricity and gasoline in Fig. 4 c.

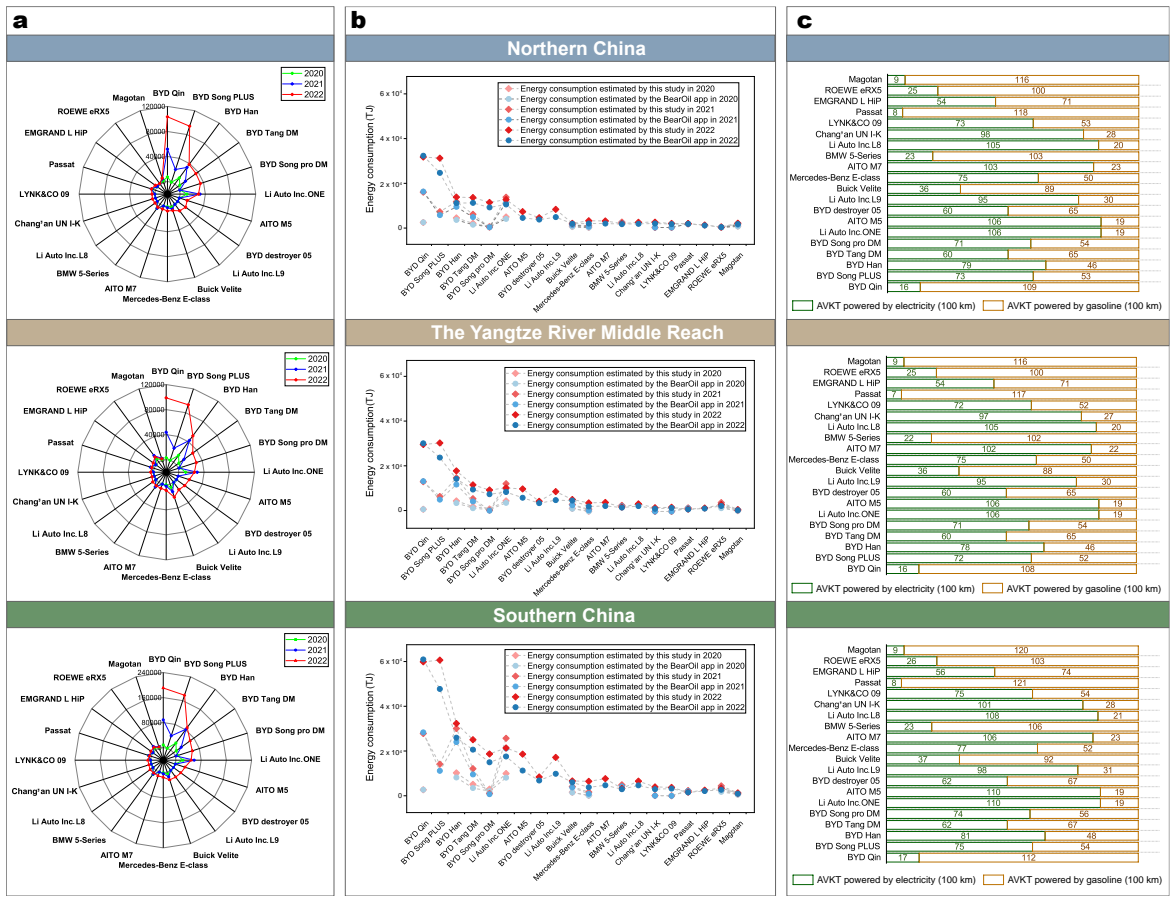

**Fig. 4.** PHEV development in different regions of China from 2020-2022: (a) trends in the top twenty selling PHEV models; (b) comparison of estimated and official total energy consumption for the operation of the top twenty selling PHEV models; (c) AVKT powered by electricity or gasoline in the operation of the top twenty selling PHEV models.

In general, the comprehensive energy consumption in Southern China was approximately double those in the other two regions for 90% of the top twenty selling PHEV models. Northern China and the Yangtze River Middle Reach exhibited relatively similar energy consumption patterns in these models. Notably, 80% of the top twenty selling PHEV models showed significant increases in energy consumption in all regions after 2020, accompanied by a noteworthy increase in vehicle sales, as shown in Fig. 4 a. Specifically, the comprehensive energy consumption of the BYD Qin increased from 455 TJ in 2020 to 4352 TJ in 2021, reflecting an approximately 9.5-fold increase. The increasing trend continued in 2022, reaching 9278 TJ, indicating a further 113% increase from 2021-2022. Ascending trends were also evident in the other PHEV models, except for the BMW 5-Series and Passat models. From a detailed perspective, the comprehensive energy consumption revealed notable variations among different models, influenced by energy intensity, vehicle sales, and corresponding AVKTs powered by electricity or gasoline. In 2020-2022, the BYD Qin had the highest energy consumption at 14,085 TJ, followed by the BYD Song Plus and BYD Han, with 7987 TJ and 7066 TJ, respectively. These results aligned with the sales of the top three selling vehicles. Moreover, extended-range EVs released in 2022, including the Li Auto Inc. ONE, L9, and L8 models and the AITO M5 and M7 models, presented relatively elevated levels of energy consumption within the top twenty PHEV models. For PHEV models derived from internal combustion engines, such as the Buick Velite, Mercedes-Benz E-Class, BMW 5-Series, Passat, ROEWE eRX5, and Magotan, as well as models released in 2021 and 2022, including the Chang'an UNI-K, LYNK&CO 09, BYD Destroyer 05, and EMGRAND L HiP, the comprehensive energy consumption remained relatively low from 2020-2022, which was attributed to fewer vehicle sales and shorter AVKTs, ranging from 315-1954 TJ.

The cumulative energy use of the top twenty selling PHEV models (as shown in Fig. 5) totaled 55,351 TJ from 2020-2022, with 39,937 TJ from gasoline and 15,414 TJ from electricity. Factors such as the increased adoption of energy-intensive technologies, economic growth, and shifts in consumer behavior significantly influenced energy consumption in all regions [57]. Consistent and notable upward trends in energy consumption were observed, with an approximately twofold increase in 2021 and a 1.5-fold increase in 2022, suggesting that the future energy demand in PHEV development will continue growing in the short term.

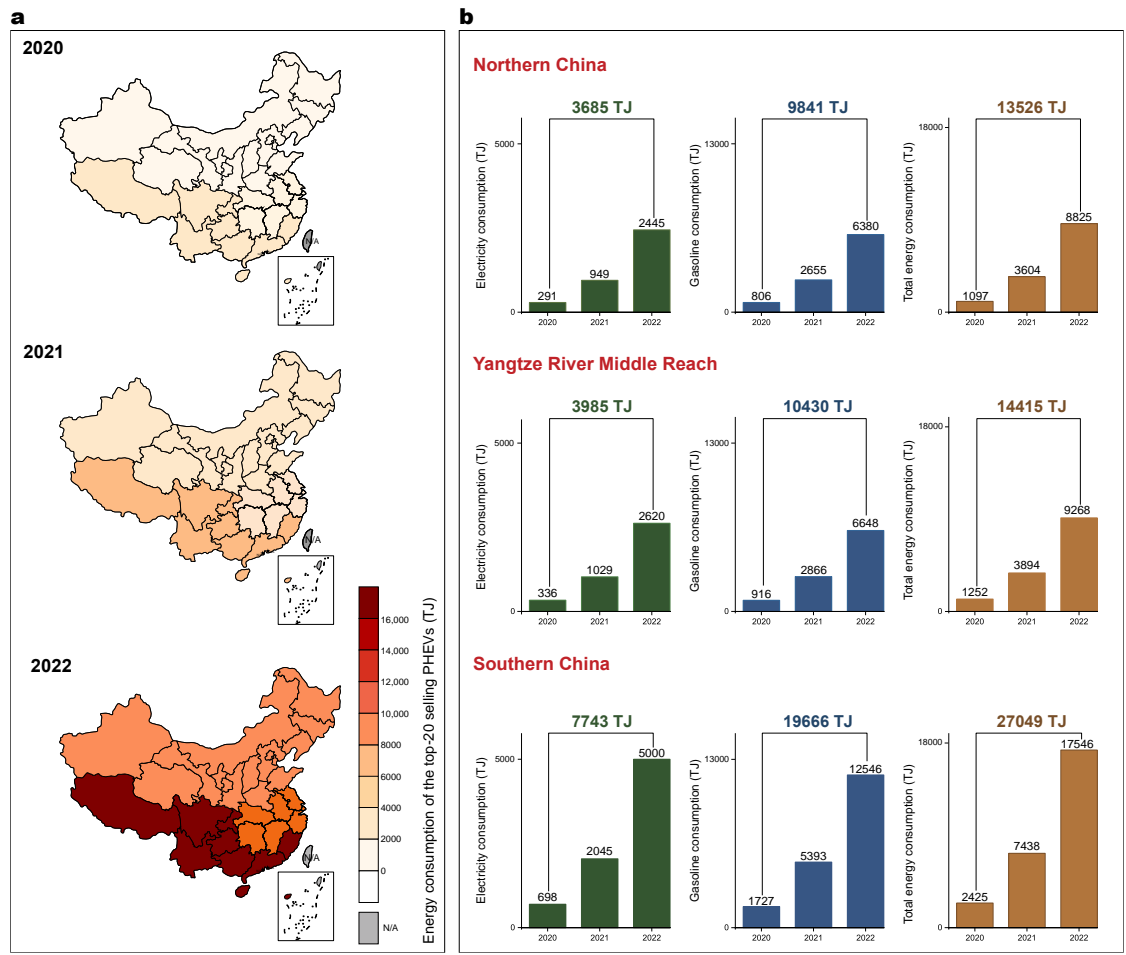

**Fig. 5.** (a) Spatial distribution of total energy consumption for the operation of the top twenty selling PHEV models; (b) electricity, gasoline, and overall energy consumption for the top twenty selling PHEV model operations among various regions from 2020-2022.

The analysis of energy consumption across different regions revealed that the top-selling PHEVs in Southern China consumed a total of 27,049 TJ from 2020-2022, which was double the energy consumption levels of Northern China (13,526 TJ) and the Yangtze River Middle Reach (14,415 TJ). This trend was consistent for both electricity and gasoline consumption, with total electricity consumption recorded at 7743 TJ in Southern China, 3985 TJ in the Yangtze River Middle Reach, and 3685 TJ in Northern China. Similarly, total gasoline consumption reached 19,666 TJ in Southern China, 10,430 TJ in the Yangtze River Middle Reach, and 9,841 TJ in Northern China. The higher consumption in Southern China can be attributed to several factors, including extensive charging infrastructure, greater reliance on PHEVs, longer AVKTs, and incentive policy measures [58, 59]. According to the 2022 annual report on electric vehicle charging infrastructure in major Chinese cities[c], the overall charging infrastructure in Southern China, with an average public charging station density of 24.3 per square km, surpassed that of Northern China, which has a density of 15.2 per square km. This led to a greater preference for PHEVs in Southern China, as evident in the sales of the top twenty PHEVs in Southern China (1,321,798 sales), which were nearly twice as high as those in Northern China (670,416 sales), and in the Yangtze River Middle Reach (722,768 sales) in 2022; moreover, the AVKT in Southern China (12,915 km) was longer than those in the other two regions. Overall, the results above summarize the total energy consumption and spatial distribution of China's best-selling PHEVs and respond to Issue 2 in Section 1.

*3.3. Operational carbon emissions from the top-selling vehicles*

Fig. 6 illustrates the operational $CO_2$ emissions, encompassing both electricity and gasoline emissions, of the top twenty selling PHEV models among various regions from 2020-2022. The distribution of $CO_2$ emissions exhibited significant disparities among the PHEV models, with notable emissions originating from high-ranking sales models such as BYD Qin, BYD Song PLUS, BYD Han, Li Auto, Inc. ONE, and BYD Tang DM, which registered $CO_2$ emissions of 1028, 708, 643, 493, and 407 kilotons of $CO_2$ ($ktCO_2$) from 2020-2022, respectively. In terms of the time period, there was a consistent upward trend in $CO_2$ emissions from 2020-2022 for most PHEV models in the three regions and nationwide, which was attributed primarily to increased sales during this period. For example, the $CO_2$ emissions of the BYD Qin increased from 34 $ktCO_2$ in 2020 to 319 $ktCO_2$ in 2021 and further increased to 676 $ktCO_2$ in 2022. From a regional perspective, the majority of the top-selling PHEV models yielded relatively high $CO_2$ emissions in Southern China, surpassing those in Northern China and the Yangtze River Middle Reach by averages of 40 and 47 $ktCO_2$, respectively. For example, PHEV models such as the BYD Qin, BYD Song PLUS, BYD Han, and BYD Tang DM were more prevalent in Southern China, with $CO_2$ emissions that were 201, 95, 124, and 63 $ktCO_2$ higher than those in Northern China. Additionally, their $CO_2$ emissions were 245, 139, 121, and 91 $ktCO_2$ higher than those in the Yangtze River Middle Reach. Conversely, 70% of the PHEV models demonstrated relatively similar total $CO_2$ emission levels in Northern

---

[c] https://tech.chinadaily.com.cn/a/202206/17/WS62abef5ea3101c3ee7adb0a9.html

China and the Yangtze River Middle Reach. The exceptions included the BYD Qin, BYD Song PLUS, and Li Auto Inc. ONE, with 44, 44, and 62 ktCO$_2$ higher emissions in Northern China, respectively, and the ROEWE eRX5, which displayed a 45 ktCO$_2$ increase in the Yangtze River Middle Reach.

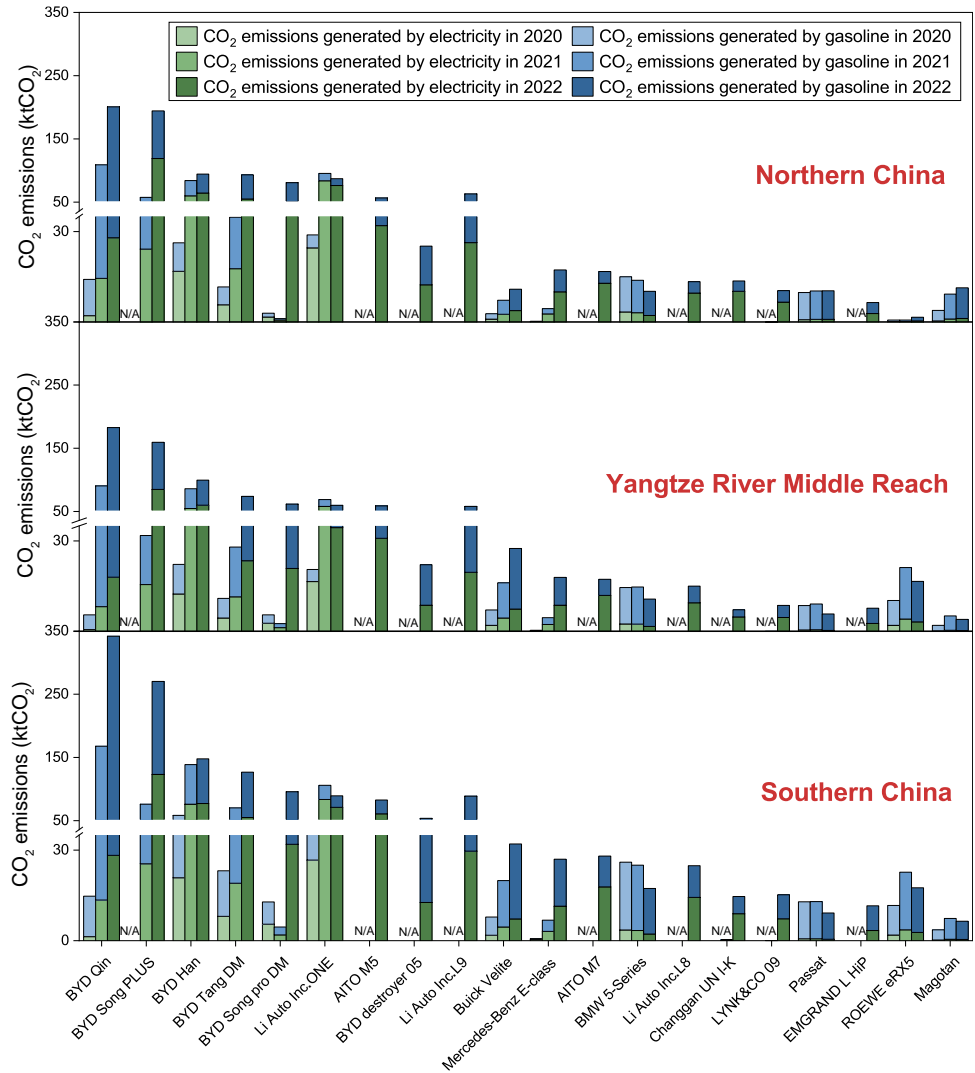

**Fig. 6.** Operational carbon emissions generated by electricity and gasoline consumption from each of the top twenty selling PHEV models among various regions from 2020-2022.

As discussed in Section 2.2, the CO$_2$ emission evaluation of PHEV operations should encompass both fuel and electricity consumption. This holistic approach is essential because PHEVs integrate a traditional internal combustion engine with a rechargeable battery and an electric motor [60]. Therefore, this study considered the carbon emission factors of both gasoline and electricity among various regions and analyzed the CO$_2$ emissions released by electricity and gasoline for the top twenty PHEV models. As shown in Fig. 6, the distributions of CO$_2$ emissions generated by electricity and gasoline for the operation of the top twenty selling PHEV models closely aligned with the electricity and gasoline AVKT determined by the real-world electricity-to-gasoline ratio (see Fig. 4 c). This shows that PHEV models with higher electricity intensity, characterized by larger BE and AER, such as the Li Auto Inc. ONE, Li Auto Inc. L8, AITO M5, AITO M7, and Chang'an UNI-K, presented relatively higher emissions generated by electricity, constituting 70% of all carbon emissions. In contrast, PHEV models with a smaller BE and lower AER, such as the BYD Qin, BMW 5-Series, Passat, ROEWE eRX5, and Magotan, are typically used as gasoline-dominated vehicles. These models had longer AVKTs powered by gasoline and tended to generate more emissions from gasoline than from electricity, accounting for 70% of all $CO_2$ emissions.

With respect to the overall operational carbon trend of PHEVs, nationwide $CO_2$ emissions from the top twenty selling PHEVs totaled 4705 $ktCO_2$ from 2020-2022. Emissions increased from 411 $ktCO_2$ in 2020 to 1269 $ktCO_2$ in 2021 and then surged to 3025 $ktCO_2$ in 2022. Combining the information illustrated in Fig. 5 a and Fig. 7 a, it can be concluded that the increasing trend in $CO_2$ emissions consistently corresponded with overall energy consumption, highlighting a significant increase over the three-year period, mainly due to the growing popularity of newly released PHEVs in the automotive market. According to our detailed analysis, Southern China, owing to the prevalence of the PHEV automobile market, exhibited significantly higher energy consumption and $CO_2$ emissions than did the other two regions. Specifically, $CO_2$ emissions in Southern China were 191 $ktCO_2$ in 2020, 567 $ktCO_2$ in 2021, and 1354 $ktCO_2$ in 2022. In contrast, emissions in Northern China were 113 $ktCO_2$ in 2020, 363 $ktCO_2$ in 2021, and 887 $ktCO_2$ in 2022, and emissions in the Yangtze River Middle Reach were 108 $ktCO_2$ in 2020, 330 $ktCO_2$ in 2021, and 784 $ktCO_2$ in 2022. Interestingly, $CO_2$ emissions in the Yangtze River Middle Reach were slightly lower than those in Northern China, despite the relatively high energy consumption in the Yangtze River Middle Reach. This phenomenon can be attributed to the comparatively lower carbon emission factors of power grids in the Yangtze River Middle Reach than those in Northern China.

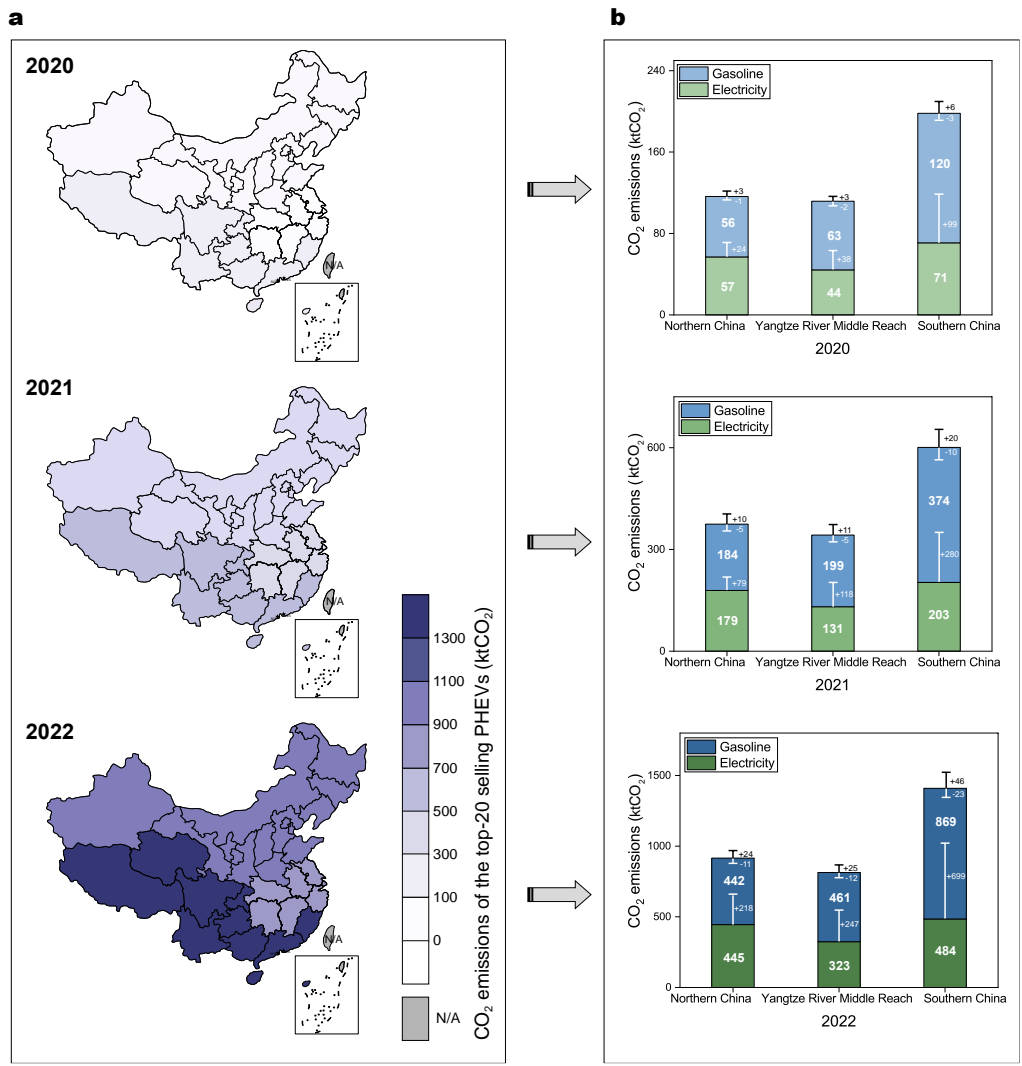

**Fig. 7.** (a) Spatial distribution of the total operational emissions from the top twenty selling PHEV models; (b) trends in carbon emissions generated by gasoline and electricity consumption from the top twenty selling PHEV models among various regions from 2020-2022.

Through detailed estimations of the energy and emissions generated separately from electricity and gasoline, as depicted in Fig. 7 b, the cumulative $CO_2$ emissions from the operation of the top-selling models nationwide amounted to 4705 $ktCO_2$ from 2020-2022. Specifically, emissions from electricity consumption contributed 1938 $ktCO_2$, whereas emissions from gasoline consumption accounted for 2767 $ktCO_2$. $CO_2$ emissions from gasoline consumption were higher than those from electricity consumption across different regions, with this trend being particularly apparent in Southern China. $CO_2$ emissions from gasoline consumption surpassed emissions from electricity consumption by 67 $ktCO_2$ in 2020, 243 $ktCO_2$ in 2021, and 519 $ktCO_2$ in 2022, suggesting that despite the increasing popularity of PHEVs, the overall environmental impact has not decreased as significantly as anticipated. This underscores the importance of continuous efforts to increase the efficiency of electric power usage and reduce reliance on traditional fuel sources [61]. Interestingly, the electricity consumption in Southern China was twice the levels in the other two regions in PHEV operations. However, the $CO_2$ emissions generated by electricity consumption in the three different regions were almost the same. This indicates that the level of low-carbon development of electric vehicles in the southern region is relatively significant. Additionally, carbon emissions per vehicle (approximately 2040 $kgCO_2$/per vehicle) in Northern China were more than 1.2 times greater than those in the Yangtze River Middle Reach (approximately 1687 $kgCO_2$/per vehicle) and Southern China (approximately 1602 $kgCO_2$/per vehicle), which was due mainly to the high carbon emission factors of power grids and the limited charging infrastructure. Overall, the above results summarize the operational carbon emission trends and spatial distribution of China's top-selling PHEV models and address Issue 3 posed in Section 1.

## 4. Discussion

Although Section 3 presents real-world energy intensity estimates for the top-selling PHEV models and highlights the significant heterogeneity in operational energy and emissions across different regions in China, divergences between these estimates and the NEDC values, as well as a more comprehensive assessment of energy and emissions, deserve further discussion. Therefore, Section 4.1 compares the estimated energy intensities with the NEDC values of the top-selling PHEV models. Section 4.2 delves into a robustness analysis of energy use estimates derived from the developed bottom-up model and the BearOil app. In Section 4.3, uncertainty and sensitivity analyses of the estimated $CO_2$ emissions in PHEV operations are conducted. Finally, Section 4.4 provides policy implications for promoting the low-carbon transition of PHEVs in the road transportation sector of China.

4.1. Comparison analysis between estimated energy intensities and official values

By comparing the estimated electricity intensities with the official electricity intensities under the NEDC conditions (see Figs. 1-2), a new finding emerges: real-world estimates consistently show a 30-40% increase over the NEDC values, despite missing official electricity intensity data for several PHEV models, especially in 2020 and 2021. The estimated values reveal that the NEDC test conditions may not accurately reflect real-world situations in China's passenger car sector and may not fully capture the complexities of everyday usage [5]. Factors such as terrain, climate, traffic patterns, road conditions, and driving habits significantly influence the real-world electricity intensity of PHEV operations [25]. Consequently, relying solely on the NEDC conditions to assess the electricity efficiency of PHEV operations can lead to misunderstandings of vehicle performance. For example, the real-world AER of the BYD Qin, reported in the *Energy Conservation and New Energy Vehicle Technology Roadmap 2.0* by the China Society of Automotive Engineers, was 54.7 km, whereas the official AER under the NEDC conditions was 80 km. Variations in the SOC, influenced by the driving mode and charging behavior, directly affect the AER values and, consequently, the estimated electricity intensity values [62]. Most PHEV drivers experience range anxiety and charge whenever a charger is available, regardless of the SOC, and continue charging until the battery is fully or almost fully charged[d], leading to a shorter real-world AER than official values [63]. Furthermore, this study used the nominal BE released by the Ministry of Industry and Information Technology, given that all of the PHEV models were within their first few years of operation. However, battery degradation over time contributes to changes in energy intensity with shorter real-world AERs [64]. Moreover, extreme temperatures, particularly low ambient temperatures, adversely impact battery system efficiency, resulting in shorter real-world AERs and higher electricity intensity [29]. Therefore, PHEV manufacturers should proactively provide consumers with real-world performance data, and the Chinese government urgently needs to introduce more realistic and reliable electricity efficiency standards to better support the development of electric vehicles nationwide.

The gasoline use intensities were estimated by the total gasoline consumption per 100 km under comprehensive road conditions, providing insight into the real-world gasoline economy of the top twenty selling PHEV models (see Fig. 3). However, a notable disparity exists between the estimated real-world gasoline intensity values and the official values under the NEDC conditions (typically approximately 3 L/100 km) provided by manufacturers and official agencies [65]. The real-world gasoline intensity was 3-6 times greater than the NEDC values, and this difference could be up to ten times greater for specific extended-range EVs, such as those from AITO and Li Auto, Inc. Essentially, the official gasoline intensities under the NEDC conditions for PHEVs were calculated based on the national standard *Test Methods for Energy Consumption of Light-Duty Hybrid Electric Vehicles* (GB/T 19753-2013) under the assumption that the vehicle operates its engine for 25 km to recharge after depleting the battery [66]. This scenario is overly idealized and significantly diverges from real-world road conditions in China, as the current civil charging infrastructure in most regions makes it challenging for most drivers to conveniently charge within 25 km [67]. However, the real-world gasoline intensity estimates of most PHEV models are close to the minimum state-of-charge gasoline intensity, except for those of specific extended-range EVs, such as those manufactured by AITO and Li Auto, Inc.

Influenced by real-world road conditions, different driving mode preferences under various road conditions (e.g., urban commuting in CD mode, highways in CS mode and other situations in blended mode), and driving behaviors (e.g., driving in a state of partial discharge, acceleration and high-speed driving, and heating and air conditioning usage) make it challenging for current PHEV models to achieve the 2025 target of 5.3 L/100 km gasoline intensity [68]. Overall, the above discussion reviews and compares the real-world electricity and gasoline energy intensity estimates with the corresponding NEDC values, providing a comprehensive response to Issue 1 posed in Section 1.

---

[d] https://theicct.org/publication/pv-china-real-world-performance-apr23/

4.2. Robustness of the bottom-up emission model for vehicle operations

In this study, the comprehensive energy consumption estimated via the bottom-up approach for PHEV operations, as proposed in Section 2.1, offers a standardized tool for cross-model comparisons. To test the robustness of the proposed bottom-up energy model, the energy consumption estimates derived from our model were compared with those calculated using the BearOil app (see Fig. 4 b). The comparison reveals a close alignment between our estimated values and the BearOil app's recorded values across different regions for the top twenty selling PHEV models, suggesting the accuracy and reliability of the proposed bottom-up framework.

Specifically, the energy use of the top twenty selling PHEV models based on our method was slightly greater than the values recorded by the BearOil app, particularly for PHEV models with higher real-world estimated energy intensity and sales [69]. This discrepancy may stem from the different estimation perspectives between our model and the BearOil app. In this study, energy consumption was estimated separately for electricity and gasoline based on separate energy use unit conversions and separate AVKTs determined by the real-world electricity-to-gasoline ratio. In contrast, the energy consumption of the BearOil app was calculated based on comprehensive energy intensity estimates and total AKVTs, without distinguishing between AVKTs powered by electricity and those powered by gasoline. However, there are limitations to our proposed model, as separate estimations may overlook the blended mode and energy-saving hybrid engine technology in some PHEV models, potentially leading to overestimated energy consumption [68]. Overall, the above robustness analysis examines the reliability of the energy use estimates of the developed bottom-up model for PHEV operations, thus completely addressing Issues 1 and 2 posed in Section 1.

*4.3. Uncertainty and sensitivity analyses*

Uncertainty in $CO_2$ emission estimates of PHEV operations arises from various factors, primarily the variability in the carbon emission factors of electricity and gasoline [6]. Detailed uncertainty ranges are shown in Fig. 7 b, and the results are also given in Appendix B. For emissions generated by electricity use, the emission factors refer to the study conducted by Zhuo et al. [54], which were generally lower than the officially released emission factors. The upper limits of $CO_2$ emission estimates were calculated using the carbon emission factors of the operating margin electricity released by the Ministry of Ecology and Environment of the PR China and the National Center for Climate Change Strategy and International Cooperation. For example, in Northern China, the upper limit of $CO_2$ emissions from electricity consumed by PHEVs could reach as high as 663 $ktCO_2$ in 2022, compared with the default value of 445 $ktCO_2$, which is an increase of 218 $ktCO_2$. For the emission factors of gasoline, this study used those recommended by the IPCC, including default, lower, and upper limits, as it is difficult to find accurate emission factors of gasoline for China. These factors account for uncertainties such as fuel properties, combustion conditions, and emission measurement techniques [51]. For example, in Northern China in 2022, the upper limit of $CO_2$ emissions from gasoline was 24 $ktCO_2$ higher than the default value, and the lower limit was 11 $ktCO_2$ lower than the default value. This trend was consistent across other regions and years, providing a comprehensive understanding of the potential variation in carbon emissions from PHEV operations.

Regarding the sensitivity analysis, variations in key model parameters, including the real-world all-electric range coefficient $\eta$, the electricity-to-gasoline ratios $\mu_{ei}$ and $\mu_{gi}$, and the electricity carbon emission factors $f_{ej}$ across different regions in China, were considered to assess their impacts on the energy and emission estimates of the top-selling PHEV operations. For example, as shown in Fig. 8, the real-world all-electric range coefficient $\eta$, initially set at 75% based on the research of Plötz et al. [53], may increase due to advancements and technological innovations in the energy system of PHEVs, leading to reduced carbon emission intensity in PHEV operations. Specifically, a 5% increase in $\eta$ (to 80%) would result in a 2.2-2.8% (39-52 $kgCO_2$/vehicle) reduction in carbon emission intensity nationwide from 2020-2022. Conversely, a 5% decrease in $\eta$ (to 70%) due to potential declines in energy efficiency, influenced by battery degradation and future vehicle engine wear, would lead to a 2.5-3.2% (45-60 kgCO$_2$/vehicle) increase in carbon emission intensity. Furthermore, variations in the electricity-to-gasoline ratios $\mu_{ei}$ and $\mu_{gi}$ of each PHEV model and the carbon emission factors of electricity $f_{ej}$ in different regions of China directly influence the carbon emissions in PHEV operations. With improvements in optimized electric systems and fully equipped charging infrastructures across different regions in the future, carbon emissions from PHEV operations will effectively decrease [34], ensuring low emissions and expediting the transportation sector's transition toward a net-zero era. Overall, the above uncertainty and sensitivity analyses of the CO$_2$ emissions estimated by the proposed model for PHEV operations completely answer Issues 1 and 3 posed in Section 1.

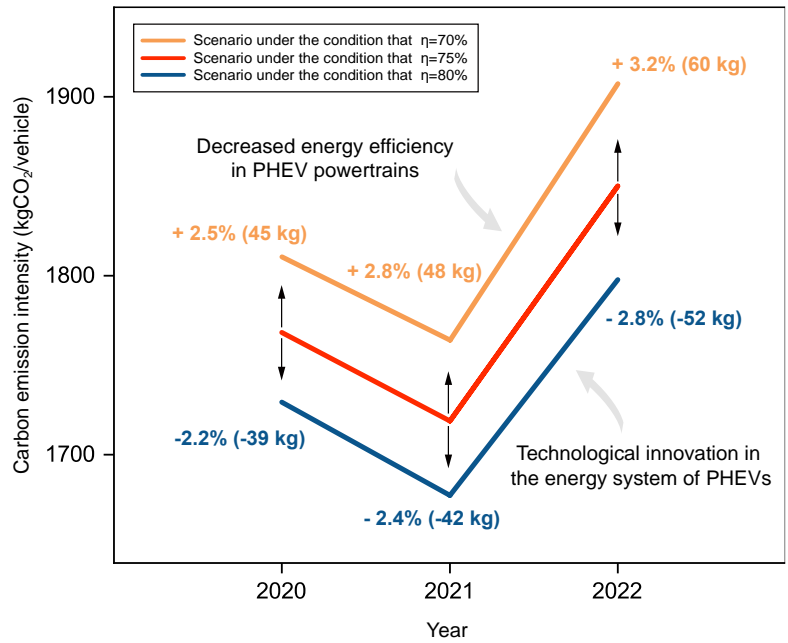

**Fig. 8.** Sensitivity analysis of the real-world all-electric range coefficient $\eta$ with a ±5% variation in carbon emission intensities of PHEV operations nationwide from 2020-2022.

*4.4. Policy implications*

PHEVs are presumed to have a development window of at least 10 years during the transitional period of electric vehicle development in China, aiming toward carbon neutrality beyond the middle of the century [70, 71]. Given several existing challenges, such as battery technology limitations and insufficient public charging infrastructure, several policy implications are highlighted for promoting the low-carbon transition in PHEV operations:

First, automobile and parts manufacturers should prioritize the application and improvement of energy efficiency and thermal control technology in PHEVs [72]. The use of sustainability assessment tools, such as exergy-based analyses, can enhance vehicle performance and adoption in the evolving landscape of sustainable transportation [73]. Additionally, PHEV manufacturers should provide consumers with performance data that reflect actual road conditions to help alleviate concerns about range and fuel consumption, thereby setting a baseline to enable dynamic modeling of future demand and emissions [74]. For better fuel economy, drivers with short commutes and access to chargers should choose battery electric vehicles or PHEV models with approximately 60 km of all-electric range, whereas those with longer commutes or less charging access should opt for extended-range EVs or fuel-efficient PHEVs derived from traditional vehicles [7]. These efforts are crucial for achieving projected gasoline consumption levels of 5.3, 4.5, and 4.0 L/100 km by 2025, 2030, and 2035, respectively, as outlined in the Energy-saving and New Energy Vehicle Technology Roadmap 2.0[e].

Given the high $CO_2$ emissions from electricity in Northern China due to its intense emission factor and less advanced charging infrastructure, it is essential to optimize the charging infrastructure in this region. Additionally, the emissions released by gasoline consistently increased and were greater than the emissions from electricity in the Yangtze River Middle Reach and Southern China. Efforts should continue to enhance the nationwide electrification process of the passenger car sector, aiming to reduce reliance on fossil fuels while continuously optimizing the clean emissions of fossil fuels [75]. This measure will promote the low-carbon transition in PHEV operations across different regions, aligning with the national goal of expediting the transportation sector's transition toward a net-zero era [76, 77].

---

[e] https://en.sae-china.org/a3967.html

## 5. Conclusions and recommendations

This work was the first to develop a bottom-up energy and emission assessment model for the operation of top-selling PHEV models among various regions in China from 2020-2022, considering variables such as PHEV model sales, real-world energy intensity estimates, AVKTs, model performance, driver behaviors, and carbon emission factors of gasoline and power grids. This study provided a robust foundation for reliable real-world data sourced from current PHEV energy demands, establishing a baseline to enable the modeling of potential energy use and corresponding emissions for PHEVs in the coming years and expediting the transportation sector's transition toward carbon neutrality. The core findings are as follows.

*5.1. Core findings*

- **The estimated electricity intensities of the top-selling models (20.2-38.2 kWh/100 km) exceeded their NEDC values by 30-40%. Additionally, the estimated gasoline intensities (4.7-23.5 L/100 km) were 3-6 times greater than their NEDC values.** Among the top twenty selling PHEV models, 60% of the samples have BEs ranging from 10-20 kWh, with AERs below 80 km. Notably, PHEV models developed on internal combustion engine platforms (e.g., the Passat and BMW 5-Series) exhibited higher fuel efficiency in terms of gasoline intensity than did recently released models such as the Chang'an UNI-K. However, influenced by road conditions and driver behaviors, the estimation of energy intensity suggests that PHEVs may not be as fuel efficient as initially anticipated. This notable discrepancy profoundly impacts the scientific accuracy of real-world carbon emission accounting. Consequently, both vehicle manufacturers and governmental bodies need to provide consumers with real-world performance data for informed decision-making.
- **The overall energy consumption of the top-selling PHEV models varied among regions: the total energy use was twice as high in Southern China (27,410 TJ, 2020-2022) than in Northern China and the Yangtze River Middle Reach.** This difference can be attributed to the greater density of charging stations in Southern China, which contributed to the increased PHEV adoption and longer AVKTs in this region. Nationally, energy use reached 55,351 TJ from 2020-2022, with a total electricity consumption of 15,414 TJ and a total gasoline consumption of 39,937 TJ. Furthermore, the robustness of the proposed bottom-up energy model was examined by comparing the results with those calculated by the BearOil app. The comparison reveals a close alignment between the estimated values and the BearOil app's recorded values across different regions for the top-selling PHEVs, suggesting the accuracy and reliability of our proposed bottom-up energy model for PHEV operations.

- **The $CO_2$ emissions from the operation of the top-selling PHEV models nationwide amounted to 4705 kt$CO_2$ from 2020-2022. Notably, emissions from electricity use contributed 1938 kt$CO_2$, whereas emissions from gasoline combustion accounted for 2767 kt$CO_2$.** In Northern China, the carbon emissions per vehicle were more than 1.2 times greater than those in other regions, mainly because of the high emission factors of power grids and the limited charging infrastructure. The top-selling models aligned emissions with their AVKTs, as determined by the electricity-to-gasoline ratio. PHEV models with higher electricity intensity and longer AERs powered by electricity emitted less $CO_2$ than gasoline-focused PHEV models. Strategically deploying PHEVs with optimized BEs and AERs customized for regional charging demands is essential for advancing sustainable development and decarbonizing the future of the road transportation sector of China.

*5.2. Future work*

This study identified several gaps that warrant further investigation, suggesting potential future research directions. One key aspect involves expanding the model samples beyond the top-selling PHEV models to include all types of electric vehicles at the provincial or city-level scale in China's passenger car sector, thereby establishing a data-driven energy and emission database to monitor the spatial-temporal future energy trend and carbon mitigation of China's electric vehicle operations. This approach will serve as a valuable tool for decarbonizing road transportation up to 2060.

**Appendix**

The appendix is included in the supplementary materials (e-component) of this submission.

**Acknowledgments**

This manuscript was authored by an author at Lawrence Berkeley National Laboratory under Contract No. DE-AC02-05CH11231 with the U.S. Department of Energy. The U.S. Government retains, and the publisher, by accepting the article for publication, acknowledges, that the U.S. Government retains a non-exclusive, paid-up, irrevocable, world-wide license to publish or reproduce the published form of this manuscript, or allows others to do so, for U.S. Government purposes.